# Real Time Monitoring and Control of Neonatal Incubator using IOT

[1]Rasha M. Abd El-Aziz, [2]Ahmed I. Taloba

[1,2]Department of Computer Science, College of Science and Arts in Qurayyat, Jouf University, Saudi Arabia

## Abstract

The care of new born babies are the most important and sensitive part of bio-medical domain. Some new born babies have a higher risk of mortality due to their gestational age or their birth weight. Most of the premature babies born on 32-37 weeks of gestation and are deceased due to their unmet need for warmth. The neonatal incubator is a device used to nourish the premature babies by providing a controlled and closed environment. This incubator provides the babies with optimum temperature, relative humidity, optimum light and appropriate level of oxygen which are same as that in the womb. But babies in the incubators have a risk of losing those babies lives due to the improper monitoring of the it which causes accidents like gas leakage and short circuits due to overheating which leads to bursting of incubators. Thus, the objective of this paper is to overcome the drawbacks of an unmonitored incubator and develops an affordable and safe device for real-time monitoring of the neonatal incubator. a low cost yet effective apparatus for monitoring the important parameters like pulse rate, temperature, humidity, gas and light of the premature baby inside an incubator. The sensed data are passed to the doctors or nurses wirelessly by the Arduino UNO via Internet of Things (IoT) so as to take necessary actions at times to maintain an appropriate environment for the safety of the lives of premature babies.

## 1. Main text

According to a study, the major reason for the 370,000 neonatal deaths in India at 2015 are the low-birth weight and premature birth. It is also found that the neonatal deaths due to the above-mentioned reasons are around 12.3/1000 births in 2000 to 14.3/1000 births in 2015. The rise in the deaths of premature babies and low weight at birth is ununiform across India. It means that the death rate is high in rural areas and less in urban areas. Furthermore, these babies needed more investments for incubators and Intensive Care Units (ICU) for getting proper neonatal care. It is found by a study that more than 20 million new born babies are premature or low-birth weight out of which 450 of them are deceased each hour. In addition to this, these babies may die due to the technical fault caused by the improper monitoring and control of the incubators.

The recent advancement in technology leads to enhancement of the medical industry and hence the mortality rate of premature babies is also controlled. This is due to the employment of incubators in the treatment of premature newborn babies. Eventhough the incubators play a vital role in the lives of premature babies, it requires instrument-health caregiver interactions due to its environmental and working conditions. Due to the ratio of number of caregivers to the number of patients is not matching i.e., more patients and less caregivers, the work load of the instrument-health caregiver is high, which leads to improper monitoring of the incubators [1]. For the sustainment of the premature babies, oxygen and nutrition are the must. In addition to this, there should be an appropriate thermal environment. Traditional incubators are reported for resulting in severe hypothermia in premature babies inside it. The thermal environment impacts the intrauterine conditions of the premature infants. The rectal temperature should be targeted to 37 to 37.5°C and also a comfort temperature is suggested. Some of the premature babies in the comfort temperature shows the rectal temperature at around 39°C. Researches are carried out to find the core temperature of the body like rectal and nasopharyngeal temperatures. The brain temperature at mild to moderate increase develops the risk of neurological damage and the mild to moderate hypothermia is considered as neuroprotective if the premature baby as hypoxia. Hence it is important to focus on the rectal and nasopharyngeal temperature in premature newborns to govern the capacity to handle the intended changes in the temperature of the incubator [2]. In Intensive Care Units (ICU), the use of central monitoring system is essential and more convenient. This helps the caretakers to monitor the patients in real-time. But this system is missing in the infant's ICU. The caretakers should go around each incubator for checking it. This makes a way to the development of web-based monitoring system after some decades [4]. A recent study shown that over 20 million new born babies are premature babies or babies with low birth weight for every year. And among them around 450 are estimated as deceased for each hour. The death of these premature or low birth weight babies can be prevented by using the neonatal incubator. The neonatal incubators are the life saving devices for many these babies by providing them with appropriate environmental and thermal conditions, so that the baby can get the normal weight and adequate nourishment by taking the available resources from the incubators [7].

The need for continuous monitoring of patients is vital in clinical environment. This must to done effectively with low cost that can solve the problem of the availability of human resources and also leaves a positive impact the life of the patients [3]. The remote healthcare monitoring is very much important for improving the quality of life of a person. The IOT makes it possible. Even the patient's personal details and which patient occupies a particular bed (Smart Beds) are also to be analyzed along with their medical and psychological details which makes the entire unit as a smart home medication dispenser and also alerts whenever the medicines are not taken in time [5]. The use of smart phones in now rapidly increasing all over the world. In addition to this, the ISP based wireless technologies with affordable cost leads to the development of many mobile applications. In 2015, it is proved that more than ¼ of population world-wide are using smart phones. The mobile applications are of social media, sports, games, news, business, healthcare and even shopping. The growth of bio-medical sensors, the need for patients and the expense of healthcare induces the developers to create an m-health application. Some of those applications are based on the staff level like sophisticated monitoring, database handling, clinical advices, diagnostic data etc. and some applications are based on the patient monitoring for vital signals, fitness, online medical advices and drugs prescriptions. The monitoring applications are divided into flat-tier and multi-tier architectures. The flat-tier architecture makes use of mobiles to collect the vital signals and passes it to the server. In Multi-tier architecture, mobiles along with a data collector are used between the sensor and the server [6].

Information technology (IT) field is developing more in the instance of sensors, nano-technology and bio-industries. The e-healthcare system is much useful in hospitals for gaining information from the patients. But these are a wired process based on the network protocol and database. Now-a-days the wireless communication is going on increasing in the healthcare field due to the advanced and new technologies built in smart devices with the help of wireless sensor nodes [8]. The use of wearable sensors is today in the next level in the field of medical diagnosis so as to help in preventing the diseases like stroke, heart attack, and so on. The need for these devices in patient's life is not only for saving the additional expenses for healthcare but also for saving the life and to enhance the life style of the patients. The major issue faced in the usage of wearable sensors are the power source. A wearable sensor should have the characteristic of long battery life. So, the development of self-powered sensors based on the body heat are proposed [9]. The wireless remote healthcare system and monitoring is usually achieved by using the wearable sensors. These devices can collect the required programmed data from the patients anywhere and these collected data are then transferred to the central storage with the help of the advanced technologies like GSM or SMS to the target mobile or computing systems of the doctor or nurse or the caretaker [11].

The use of the incubator is to maintain the adequate environment for the premature new-born babies. This is because the premature infants have limited immunity and limited thermoregulation. They are more sensitive to the environmental conditions. Even a small change in the surrounding can cause adverse effect in them. Hence, there is a need for some artificial device to make those infants sustain in the world. It is reported that around 1.49 million oof premature babies with low birth-weight are born in 2010 all over the world. Moreover, the distance of the hospitals are almost 8 kilometers away from the patient's residency even in the developing countries [10]. In bio-medical field, the most complicated part to be focused is the premature infants. Some of the premature infants are at a higher risk than the average risk, due to the gestational age or the birth weight that risks the life of those babies from diseases or death. The infants in Neonatal Intensive Care Unit (NICU) are mostly preterm and the risk associated with them is due to the prematurity. A statistical analysis from Iran in 1980 reported that, 13% of newborns were preterm and in 2006 it rises to more than 30%. The environment of the preterm babies should be the same as that in the womb so as to survive for them. The recent studies showed that each preterm baby has a microenvironment which is based on the gestational age and medical condition of them. Hence the incubator is used here, as it can fulfil the need for the microenvironment in the way of stable temperature, humidity, oxygen and light conditions so as to protect those infants from getting infected by the surroundings and also to sustain in the physical environment until they are completely cured to be survived. The air temperature should be 35ºC [12]. All over the world, around 1 million of preterm babies or low birth weight babies die on the first day itself and 4 million of them die in the first month. Almost a minimum of 25% neonatal deaths are associated directly or indirectly with the preterm birth and the infants with low birth weight are at high risks. Around 1.8 million infants die every year due to improper maintenance of the temperature in their body [13].

The World Health Organization (WHO) reported that, globally more than 1 of 10 pregnancies results in preterm birth, which means a birth that happens after the pregnancy but not more than 37 weeks, as the normal pregnancy should be of 40 weeks. the preterm birth can be classified into 3 stages based on their gestational age. They are the (1) Late preterm ie., 32-37 weeks, (2) Very Preterm ie., 28-32 weeks and (3) extremely preterm ie., < 28 weeks. The preterm birth is the major reason for the death of the children below 5 years and the mortality rate of them increase every year in underdeveloped environments. The preterm birth can be prevented by providing steroid injections or antibiotics and essential care for the newborns. But the use of incubators reduces the risk of premature babies by 24%, as the incubator can provide the essential environmental conditions for the newborns to be survived [14]. The incubator makes the body temperature of the infant to be stable inside a controlled and sterile environment. In case of infant mortality rate, the incubator is a precious device for the infants with prematurity. For example, the Malawi is the highest of the rate of preterm births worldwide. In addition to this, the incubators are more expensive which cannot be affordable for countries under development. Also, the power needed for the incubators are also high. Moreover, there will be a difficulty in the transportation of such a complex machine across the rural areas. Hence the need for a portable, yet durable and alternative powered incubators are much essential for the developing countries [15]. The Ballistocardiograph (BCGs) is a device that captures the heart's mechanical activities. The researches related to the BCG are going on due to its benefits in recording unobtrusive psychological measurements. The BCG can also monitor the bio-signals of infants without any physical confinement [16].

The wired communications of biological data are to be replace with the wireless sensors so as to make the work of the healthcare professionals to be easier. This method is to needed to be applied for the wireless and automated monitoring for the infant's heart rate, body temperature and movement every minute for the early diagnosis of the risk of being affected with cardiac problem, abnormal boy temperature and body movement, hypothermia and hyperthermia and act accordingly for early recovery and survival of the infants [17]. The screening of infant health is an essential part of the public health that comprises of neonatal screening, monitoring the nutritional status and diagnosis of infectious disease. But it is more complex for the caretaker due to the crude nature and the need for manual inputs of the machines and the lack of training for the duty staffs that leads to error and falsification in the results [18]. The premature babies below 37 weeks of gestation age has a low body weight and are categorized as (1) Low Birth Weight < 2500g and (2) Very Low Birth Weight < 1500g. Also, their body parts or organs are not yet developed completely. So, these babies are needed to be taken care more seriously to survive by making an intensive care within the incubator immediately for neonatal assessment. In 2010, Indonesia is one of the highest among the countries with premature birth rates. It is found that premature birth is more common in the rural area that are with less healthcare facilities and in the families with less income. The NICU is the place where the premature infants are treated with the incubator. But the NICU is available only in the hospitals in big-cities also it is more expensive that cannot be affordable for low-income families and families in rural areas [19]. The thermo-regulation is a key issue in the premature babies that is to be taken care of the most. Often, they are kept inside the incubators with convective heating. The air temperature in the incubator is sensed which is then used for controlling the flow of heat or the skin temperature is sensed which is used in closed loop control. But the skin temperature often results in huge fluctuation in the air temperature inside the incubator and the air temperature causes fluctuations in the skin temperature. So, it is a big question mark whether the air temperature and the skin temperature can be controlled simultaneously [20].

## 2. Related Works

An incubator with a wireless transmission of alerts to the neonatal nursing station was developed. This system can minimize the workload of the caregiver. This system is categorized into 4, namely the incubator temperature monitor and control unit, body temperature monitoring unit bed wet monitoring unit and alarm transmission. The temperature is measured by using the LM35 temperature sensor that is fixed within the incubator and then this signal is read by using the USB4704 and controlled by LabVIEW. The control signals are generated by using the ON-OFF control logic was used and these signals are generated via the USB4704. Also, the body temperature was read using the same USB4704. The detection of bed wet was done by a 5V power supply-based circuit. Finally, the alarm signal for sensing any unfair condition of these circuits is transmitted to the remote nursing station [1]. The premature babies are investigated for their ability to cope with the temperature change with their rectal and nasopharyngeal temperatures heat flux in several site and heart rate by either increasing or decreasing the temperature of the incubator to 1°C. these values are to be recorded for every 6 hours. After that it is found from the investigation that the rectal, nasopharyngeal and skin temperatures are significantly increased when the temperature of the incubator is increased to 1°C when compared with the 1°C decrease in temperature. Hence it is evident that the premature infants have less capability to cope with the temperature increase in the nasopharyngeal and rectal temperatures [2].

A prototype of a low-cost modular monitoring system design that aims to support the mobile devices to provide a better and yet fast medical interference in emergency situations is introduced. This system is developed by using the low-power sensor arrays for EKG, SpO2, temperature and movement which are interfaced based on IoT. The control unit uses RESTful web interface for safe and flexible integration of the objects [3]. The remote healthcare technology reduces the need for managing the chronic diseases regularly with the patient's intervention also it enhances the quality of those people. But due to the emergence of the Internet of Things and the low-cost sensors, the offline maintenance of all the details needed in the monitoring during the runtime by engineers can be drastically reduced. Hence, a system for ad-hoc based healthcare monitoring with low-cost wireless sensors with the IoT technology was built. This prototype of a basic remote healthcare monitoring system monitors the patients real-time and alerts the patient's caretakers or doctors whenever the system experiences an unusual condition that requires any medical consultation [5].

The incubator which can be operated, managed and monitored in real-time for inspecting the humidity and temperature of the incubator by a web-based system with the help of the Intranet and this type of system is installed into the Intensive Care Unit (ICU). A pilot system is developed with the temperature and humidity sensor and a module for measuring it in each of the incubator that are connected into the web-server through Rs485 port. This model transfers the signal via the standard TCP/IP protocol so as to accessed by any of the users with the computers with internet inside of the healthcare center. This signal provides us with the humidity and temperature values of the incubator measured by the modules through RS485 port on the web server to create a web-based document with those gathered data. This system allows the hospital staffs, a centralized monitoring of all the incubator conditions inside of the ICU with a computer. Also, this system alerts the staffs by giving an alarm signal when the incubator experiences an unusual behavior when exposed to the light or sound with the help of the measuring module integrated with each incubator. If this centralized supervisory monitoring station is integrated with many of the incubators, then this system can be much convenient for the staffs in the ICU for better intervention of the incidents happening in the incubators inside the ICU [4]. There are a greater number of preterm babies and is increasing all over the world. M-health is a application that plays an important role in the monitoring of bio-medical parameters with the help of internet in the mobile devices. Hence, a Distributed Neonatal Incubator Monitoring System (NIMS) was developed for the monitoring of premature infants. This system is built by using distributed software agents inside the data hub of the incubator, medical server and the mobile terminal of medicals staffs and parents with the help of Constraint Application Protocol (CoAP). The agents of CoAP and the URL for Data ID are used for to enable integrating the NIMS and data collections to IoT, and this system was tested in North Lebanon hospital [6]. A cost-worthy and real-time monitoring system for the premature infants inside an incubator was investigated and designed based on the embedded system. This system focuses on the diagnosis of the life-threatening actions in the earlier stage for maintaining a safer environment needed for the babies. Both of the small and medium companies based on medical technology mostly doesn't adopt the existing best technologies, since they are all most expensive. Though, the big medical companies can afford it, common man can't. Thus, a cost-effective yet eco-friendly monitoring system that can be easily affordable by a common man is proposed in [7].

RFID (Radio Frequency Identification), WSN (Wireless Sensor Nodes), etc. are used for the identification and information processing of any devices. The Body Area Network (BAN) is implanted into the body. Today's e-healthcare applications are connected with the smart phones, tablets or computers for the transmission of the data collected from the patients related to their health. Hence, a smart phone based patient health care monitoring for transmitting the patient's health related data to the medical staffs through IoT. This system monitors the patient's health parameters regularly and compares it with the pre-defined parameters and whenever the real-time data varies from the pre-defined data, then it passes an alert signal to the medical staffs to take proper actions. The (Wireless Body Area Network) WBAN when associated with the smart phones, develops vast practicality. Hence this system is supposed to be more capable worldwide for e-healthcare [8]. The wearable healthcare sensors are expected to have the capability of changing the healthcare system in today's world. There are many advanced technologies being introduced with enhanced performance, but the major concern still to be taken care of, is the battery runtime for the use in practical life. The runtime of a battery is limited to the size of the battery, and it is the next major concern in practicality and acceptance from the customer. Hence, a new system Ultra Low Power Sensor Evaluation Kit (ULPSEK) was introduced for evaluating the sensors and the application (http://ulpsek.com). This system consists of multi-parameter sensor for the measurement and processing of ECG, motion, temperature, respiration and PPG (Photoplethysmography). The ULPSEK system can be powered by the body heat harvester instead of using a battery for power. An average of about 171 μW power is harvested from the ULPSEK that is enough for the power supply of sensors used at an ambient temperature of less than 25°C. By keeping in mind, the security issues arise from the self-powered sensors, a hybrid solution was suggested by using a battery charged by the harvester [9]. By the guidelines provided from WHO and Engineering World Health (EWH), a prototype was designed which modifies the existing HVAC system for the optimization of the essential components such as the Humidity Control and Sensing, Heating System and the Quantitive Thermal Optimization and tested it and obtained a safe and efficient design for using wherever needed. In this redesign, the aluminum piping was replaced by the polyethylene for efficient flow of air with the help of the computer fans. This replacement is done because, the polyethylene provides a better modularity and can be obtained non-corrugated, so that the resistance to airflow can be reduced and also the bacterial growth from trapped moisture can be reduced [10].

A prototype was designed based on wireless healthcare monitoring by sending SMS of the healthcare paramaters to the caretakers. This design process is categorized into 3 phases. the data acquisition phase, the data processing phase and the communication phase. In the phase of data acquisition, the temperature and pulse rate sensors are used to monitor the temperature and the pulse rate of the patient. This output is then converted into digital format and is passed to the Basic Stamp microcontroller for processing before sending it to the data processing software called Visual BasicTM via a serial port. A GUI form related to patient's data was created using the Visual Basic for assisting the data processing step. In this step, some values can be pre-set for the essential parameters of medical sensor data. Whenever the system finds the exceed in the pre-set values, immediately a communication will be initiated by sending an SMS to the uploaded mobile number. When unknown values are detected by the sensors, then the communication will be by sending the SMS to the doctor's mobile phone through the GSM or SMS gateway, simultaneously the database from the hospital will also be updated periodically. Several different designs are made and chosen the best from them. The data acquisition and accuracy of the sensors are somewhat compromised by using some available sensors instead of bio-medical sensor, so as to reduce the additional cost and for the availability factor [11]. A closed-loop control system was designed and is implemented for regulating the humidity, temperature to maintain the optimum temperature of the body and the incubator and the intensity of light for preventing from jaundice and oxygen within the incubator. When the sensors combined with the actuators, a closed loop control system will be formed. They are operated synchronously for giving a stable thermal environment within the incubator. The implementation of this system was done by using the microcontroller and PID controller [12]. An incubator monitoring system with a single-chip microcontroller was integrated with the traditional incubator monitoring system. A temperature sensor for sensing the temperature in the incubator, a humidity sensor for reading the humidity value in the incubator and a respiration sensor for reading the respiration rate of the infant in the incubator, pressure sensor, cooling fan, LCD and SIM SMS MODEM are connected to the proposed model of incubator monitoring system. This single-chip microcontroller reads and frames the parameters like surrounding humidity, temperature and respiration in addition to the sensors in the incubator. These recorded values are displayed in the LCD. A small variation in the sensed data when compared to the pre-set values, immediately an alert will be sent to the parents of the infant inside the incubator automatically with the help of the single-chip microcontroller [13].

A premature infant's birth is indirectly connected to some serious care that are to be taken for the survival of those infants by monitoring the movement, weight gain, etc. of the infant. Along with these, other parameters like the temperature and humidity of the incubator environment are also to be monitored to prevent loss of heat in the infants. Hence, a modern incubator with temperature and humidity sensor and weight sensor was demonstrated. This incubator is then connected to a central network based on Long Range Networks (LoRa) which is used for storing the medical data in a database. Also, in this system, a Near Field Communication (NFC) interface was introduced which can be used to identify the doctor who is giving the treatment, the patient's evolution and if any updates in the database done by the doctor. At last, this system is subjected to a laboratory test for verifying its functioning [14]. A structural design of an incubator focusses on the durability and the portability. The base of this incubator is made of the luggage that is converted based on the requirement. The vestibule is made of a pop-up tent design which can be collapsible and this cover was made of a clear and washable plastic which can support 2 DC heaters that was powered by the main-line power or by a solar rechargeable battery. These heaters can maintain a steady temperature ranges from 34°C to 37°C inside the incubator. This model was expected to be improve the lives of the premature infants and reduces the mortality rates [15]. A psychological signal monitoring bed which is based on load-cells was designed and described. An algorithm is subjected to extract the breathing rate and the heart rate of the patient by the signals obtained from the load-cells. An experiment was carried out to test this system by using 13 infants. The ECG and the respiration signals are also monitored by commercial devices for comparing with the generated signal for accuracy. Optimal respiration and heart beat sensors are connected to this algorithm. The results obtained form the load-cell sensor shows an average error or 2.55% in heart beat and 2.66% in respiration performances. Also, in this experiment, the positive feasibility of the BCG-based measurements is verified in the babies [16]. An automatic neonatal monitoring system was designed by using the sensor mobile cloud computing which is based on the wireless sensor network and the mobile cloud computing. This system consists of a temperature sensor for monitoring the temperature of the neonate, an acceleration sensor for monitoring the acceleration caused by the movement of the baby, and a heart rate sensor for measuring the heart rate of the baby. The body temperature of the infant was monitored and controlled by using the wireless radiant warmer. The readings obtained from the sensors are uploaded to the cloud. The neonates are continuously monitored by the caretakers by accessing the sensed data via an android application in their mobile phone. And whenever an abnormal condition is happening in the system, then an alert will be given to the caretaker's mobile phone automatically and hence the required care would be provided to the needed neonate in the early stage which leads to early recovery [17].

A design of a low-cost child assessment kit named Baby Naapp was demonstrated with a smartphone using less expensive peripherals. This system is then subjected to a laboratory testing and proved that it compares favorably against the traditional manual measurements. In addition to the basic tools on measuring the anthropometric measurements, advanced tools are also demonstrated for photoplethysmography and thermal imaging that can be helpful for the community healthcare staffs for screening and diagnosis in the primary care clinics. Also, the measurement of bilirubin for jaundice and hemoglobin for anemia are suggested to be included with the Baby Naapp toolkit [18]. A remote monitoring system of the preterm baby's weight within an incubator was designed using the Arduino UNO R3 microcontroller, load cell as a sensor and active filter. This system is completely integrated with a GSM module named SIM900 for transmitting the acquired data of the neonate's body weight to the medical professionals. This system is then examined by making use of a moving baby doll for the simulation of moving baby in an incubator. The result showed that, the outcome of this system is stable even when there is a motion of the baby inside the incubator, also this system can be able to transmit the captured data to the mobile device through SMS [19]. A fuzzy logic system was proposed for controlling the heating problem occurring the incubator by monitoring the temperature of both the air and skin of the babies inside the incubator. This system was examined with a lumped parameter model of infant-incubator system and is proved that the on-off skin control causes variations the air temperature inside the incubator, resulting that the air control causes slower rise-time of the core temperature. Thus, the fuzzy logic system can provide a smoother control with desired rise-time [20].

## 3. Proposed Methodology

Whenever the sensors are combined with IoT technology, it can be an illustration of a cyber-physical system, smart home, smart grid, smart city, intelligent transportation and virtual power plants. Therefore, the IoT can help in controlling the objects remotely being sensed by the sensors so that the physical systems can be easily integrated by the computing systems. This advantage not only improves the accuracy and efficiency of a machine but also minimizes the human intervention needed for monitoring a machine or a device 24/7. The use of ThinkSpeak platform makes easy to access the data stored into it by means of IoT-Data Analytics. Moreover, the used sensors in the proposed models are of low cost and easily affordable so that the cost of periodic maintenance is also easier.

The proposed system consists of an Arduino UNO microcontroller, which is to be connected directly to the incubator and several sensors are used to sense the biological signals inside the incubator and in the body of the premature infant. The various sensors used in the proposed model of neonatal incubator for real-time monitoring and control includes, the Temperature and Humidity sensor (DHT11) for sensing the temperature and the humidity in the surroundings of the neonate, the Pulse rate sensor to record the heart rate of the infant, the Gas Sensor to sense the additional gas leakage and the Light sensor to capture the extra light penetration. Additionally, the IoT Module (ESP8266) is used to for transmitting the recorded or sensed data wirelessly which are to be uploaded in the ThinkSpeak platform, which is an open source IoT and Application Programming Interface that is used for the purpose of storing and retrieving the data from sensors which are then transferred to the receiver's device and an LCD is also connected to the microcontroller to display the recorded signals near the incubator for monitoring.

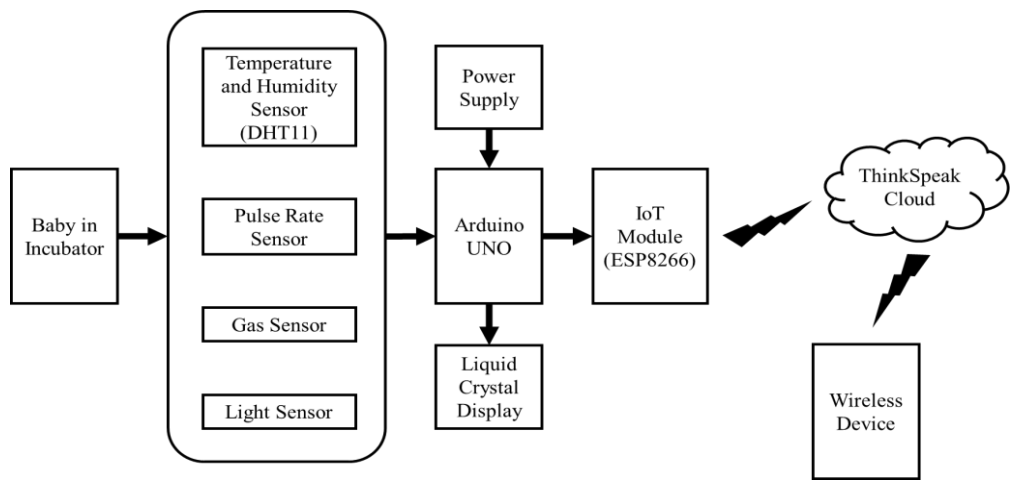

Fig. 2. System architecture.

The proposed neonatal incubator design consists of 3 sections namely, the terminal device, the network protocols and the monitoring and control of the neonates. The various sensors connected are used for monitoring and controlling the incubator via IoT. The Arduino microcontroller is programmed in such a way to get the output of these sensor and display it on the LCD for monitoring purpose. The values of the sensors are then uploaded in the ThinkSpeak cloud to display it on the wireless device in the receiver side, which is then used for enabling the control of the incubator environment.

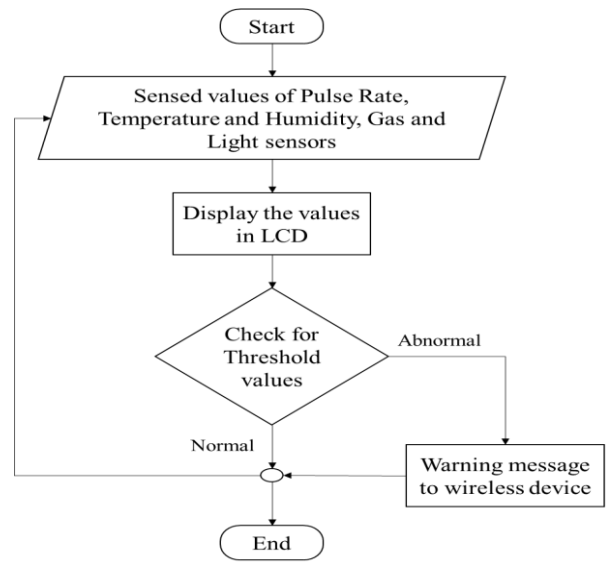

Fig. 2. Flow chart.

The DHT11 humidity and temperature sensor reads the humidity and the temperature in the surroundings of the incubator environment. The optimum value of temperature should range from 36.5°C to 37.2°C. If the temperature value exceeds the preferred range, then the proposed neonatal incubator system alerts the caretaker of the neonate in the incubator via IoT to take necessary actions. Similarly, the light sensor detects the intensity of light penetrating into the incubator and the gas sensor detects if there is any gas leakage inside the incubator and if the value of light and gas exceeds the optimum range, then the system alerts the caretaker regarding this through IoT. Also, the pulse sensor monitors the heartbeat of the neonate and the same alerting system is being executed whenever necessary.

## 4. Conclusion

The proposed real-time monitoring and control based neonatal incubator monitors and detects any changes in the environment surrounding the incubator like pulse rate, temperature, humidity, light and gas values with the help of the pulse rate sensor, temperature and humidity sensor, light sensor and gas sensor respectively and sends those signals to the microcontroller, Arduino UNO and the controller then alerts the doctors or nurses or the caretakers of the neonates inside the incubator via IoT to take necessary and possible actions so as to maintain the health of the preterm infants within the incubator. This system helps in preventing the unusual accidents and deaths that occurs in an incubator due to improper monitoring of it. Eventhough the proposed neonatal incubator monitors all the essential parameters needed for the environment of a preterm baby, there is still an issue of exposure to high level of noise inside the Neonatal Intensive Care Unit (NICU). Hence the behavior of NICU should be modified so as to reduce the noise exposure. In addition to this, the Electromagnetic Fields (EMF) impacts the health of the preterm baby is still unclear. Therefore, in future the preterm babies' incubators would be designed in such a way that the noise and EMF exposure to the neonate to be minimized.